\documentclass[a4paper,12pt]{ETHpaper}

\usepackage{cite}
\usepackage{amsmath,amssymb,amsfonts}
\usepackage{algorithmic}
\usepackage{graphicx}
\usepackage{textcomp}
\usepackage{xcolor}
\def\BibTeX{{\rm B\kern-.05em{\sc i\kern-.025em b}\kern-.08em
    T\kern-.1667em\lower.7ex\hbox{E}\kern-.125emX}}
\usepackage[numbers,sort&compress]{natbib}
\usepackage{caption}
\usepackage{tabularx, dcolumn, booktabs}
\usepackage{subfigure}

\begin{document}
\renewcommand*{\thefootnote}{\fnsymbol{footnote}}
\title{Quantifying Triadic Closure in\\ Multi-Edge Social Networks}
\titlealternative{Quantifying Triadic Closure in Multi-Edge Social Networks}

\author{Laurence Brandenberger\footnote{lbrandenberger@ethz.ch}, Giona Casiraghi\footnote{gcasiraghi@ethz.ch},\\ Vahan Nanumyan\footnote{vnanumyan@ethz.ch}, Frank Schweitzer\footnote{fschweitzer@ethz.ch}}
\authoralternative{L. Brandenberger, G. Casiraghi, V. Nanumyan, F. Schweitzer}
\address{Chair of Systems Design, ETH Zurich, \\
Weinbergstrasse 56/58, 8092 Zurich, Switzerland \\[2mm]
}
\www{\url{http://www.sg.ethz.ch}}

\maketitle
\renewcommand*{\thefootnote}{\arabic{footnote}}\setcounter{footnote}{0}
\begin{abstract}
Multi-edge networks capture repeated interactions between individuals.
In social networks, such edges often form closed triangles, or triads.
Standard approaches to measure this triadic closure, however, fail for multi-edge networks, because they do not consider that triads can be formed by edges of different multiplicity.
We propose a novel measure of triadic closure for multi-edge networks of social interactions based on a shared partner statistic.
We demonstrate that our operalization is able to detect meaningful closure in synthetic and empirical multi-edge networks, where common approaches fail.
This is a cornerstone in driving inferential network analyses from the analysis of binary networks towards the analyses of multi-edge and weighted networks, which offer a more realistic representation of social interactions and relations.

\vspace{1em}
	\textbf{Keywords}: multi-edge networks, triadic closure, network inference, social networks, statistical learning
\end{abstract}

\section{Introduction}

\noindent Triadic closure denotes an important tendency observed in social networks, namely to form triangles, or triads, between three individuals $a$, $b$ and $c$ (see Fig.~\ref{fig_triangle}).
That means, if $a$ and $c$, and $b$ and $c$ are connected by an edge, there is a larger probability that also $a$ and $b$ will form a edge.
A large number of empirical studies have reported significant effects of triadic closure in social networks \cite[for an overview, see][]{rivera2010dynamics}.
Triadic closure is closely related to other relational mechanisms such as  reciprocity \cite[see for instance][]{block2015reciprocity} and assortative mixing \cite[see for instance][]{feld1982patterns, goodreau2009birds}.

The presence of closed triads can be quantified on the topological level, for example by calculating the clustering coefficient.
However, often our interest is in examining whether triadic closure is an actual mechanism driving network formation, or whether observed triads are simply a consequence of the other relational mechanisms of the network.
Similarly, it is important to know to what extent other network properties of interest, e.g. communities or core-periphery structures, are already determined by triadic closure.
For these reasons, triadic closure is often used as a control variable in inferential network models.
This, however, requires that triadic closure can be included correctly in such inferential models.
In our paper, we solve this problem for cases which are important for real world applications, but have not been appropriately addressed so far.

Triadic closure is well-defined only for binary networks, i.e., networks with either present (=1) or absent (=0) edges. 
In such networks, measuring triadic closure can be operationalized by counting triads.
Most real networks, however, reflect \emph{repeated interactions} between individuals, which are quantified by multi-edges, i.e., each interaction is represented by a separate edge and there are multiple edges incident to the same pair of nodes \cite[][]{bollobas1998modern}.
Alternatively, instead of multi-edges, weighted networks are used in which the weight represents the number of interactions.
For those networks, however, an operationalization by counting triads faces two problems.
First, it neglects important information, because the number of triads does not appropriately reflect the possibly different interaction intensities among two nodes, i.e. the fact that within a triad nodes can repeatedly interact a different number of times.
Second, because of these repeated interactions, multi-edge networks often have a high density, as measured by the number of edges divided by the number of nodes.
This would result in a maximum triad count for the network, overrating the fact that some of the nodes only have interacted once.

In this paper, we present a way of testing for triadic closure in repeated interaction networks which overcomes the mentioned problems. 
In short, our operationalization focuses at the dyad level, i.e. on the relations between two nodes $(a,b)$.
First, we count the number $k$ of shared partners a dyad $(a,b)$ has. 
The two edges each of these partners $i\in k$ has with $a$ and $b$ form a \emph{two-path} ($a, i$)-($b, i$), i.e. a path from $a$ to $b$ that goes through $i$. 
Further, these partners can have multiple interactions with the nodes $a$ or $b$, reflected as multi-edges.
\emph{Edge counts} (also referred to as edge multiplicity or edge weights) $v(a,i)$, $v(b,i)$ then give us the number of these interactions. 
For our calculation, we take the minimum edge count between edges ($a, i$) and ($b, i$).\footnote{Even though we present an undirected version of triadic closure in this article, our approach can easily be generalized to operationalize different directed versions of closure (i.e., transitive triplets or cycles).}
This ensures that triangles involving two-paths ($a, i$)-($b, i$) with very small edge counts are given less weight than
triangles involving two-paths with many interactions among $a$, $b$ and and $i$.

To test our operationalization for multi-edge networks, we first validate our approach using synthetic network data. 
Second, we use two different classes of inferential models, 
(i) Exponential Random Graph Models (ERGM) for count data \cite[][]{krivitsky2012exponential} and (ii) generalized hypergeometric ensembles (gHypEG) \cite[][]{casiraghi2016generalized, casiraghi2017multiplex} to measure triadic closure in two real-world data examples.
Our results show that triadic closure is correctly inferred even for the case of multi-edge networks.

\section{Theory and Operationalization of Triadic Closure in Social Networks}

\subsection{Triadic Closure in Social Networks}

\noindent Closure is a key feature of most social networks.
In its simplest form, closure refers to a relational mechanism where three nodes connect to form a closed structure, a so-called triangle, or triad.
But how do these clustered connections develop over time?
The main theoretic rationale behind (triadic) closure is that established connections in social networks create opportunities to meet and interact with new nodes, and develop new connections (i.e., edges).
Reference \citep{simmel1950sociology} developed this argument by noting that such opportunities to interact can stem from joint group memberships and that they often lead to clustering effects in larger social groups.
Reference \citep[][]{granovetter1973strength} builds on this argument and theorizes that people are more likely to get to know the friends of their friends due to increased opportunities to interact and share time together.

Apart from increased opportunities to interact, repeated interactions with close friends of friends also allow for a build up of trust among third parties \cite[][]{granovetter1985economic, coleman1990foundations}. 
Reference \citep[][]{burt1995kinds} argue that indirect connections can ease access to information on third parties, and thus facilitate interactions between two previously unconnected people by reducing risks of engaging with them. 
Thus, people have a tendency to interact positively with friends of their friends and eventually are more likely to consider them friends as well.
Whenever there is hostility among unconnected third parties, however, it creates an imbalance in the triad, which needs to be resolved. According to Balance theory, postulated by \citep[][]{heider1946attitudes}, social relationships have to be balanced in order to be long-lasting. Unbalanced relationships are difficult to maintain as they cause discomfort and put a strain on relationships \cite[][]{newcomb1961acquaintance}. 
For instance if node $a$ is friend with $b$ but dislikes $c$, it is difficult for $b$ and $c$ to become friends as their mutual acquaintance $a$ is both praised (by $b$) and despised (by $c$) and thus acts as a source of discomfort in the interactions between $b$ and $c$.

Friendship networks, and especially adolescent friendship networks, have strong tendencies for closed triads in the network \cite[see e.g.,][]{rambaran2015development}. Evidence of triadic closure has been shown in political networks \cite[e.g.,][]{lerner2013modeling}, collaboration networks \cite[e.g.,][]{newman2001random}, or communication networks \cite[e.g.,][]{kossinets2006empirical}. 
Reference \citep[][]{rambaran2015development} examine friendship formation in two U.S. middle schools (480 nodes) over the course of three years and give longitudinal evidence that adolescents form friendship ties in a balanced way and that mutual dislike of a third party increases a friendship bond between two adolescents.
Reference \citep[][]{leifeld2012information} examine an information exchange network among political actors involved in drafting a new policy against chemical pollution in Germany in 1985 (30 nodes) and show that by controlling for triadic closure in their network, belief similarities (which have always been assumed to be the building blocks of policy networks) no longer explain information exchange patterns.

\subsection{Closing Triads in Multi-Edge Networks}

\begin{figure}
\begin{center}
	\includegraphics[width=1\columnwidth]{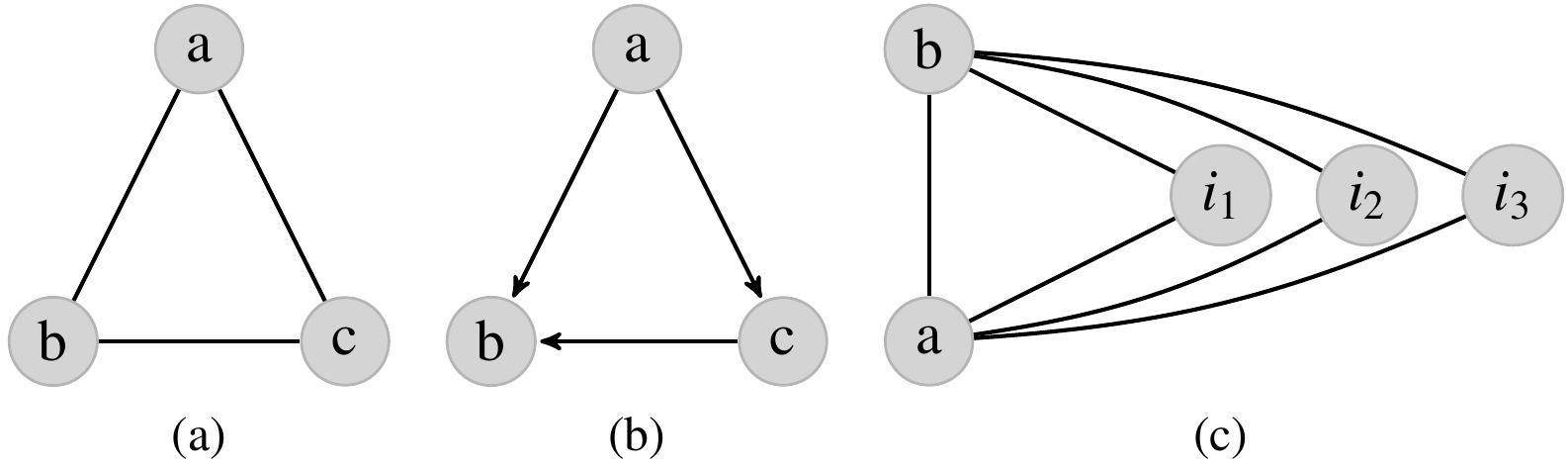}
\end{center}
\caption{Triadic closure: (a) undirected triangle, (b) transitive triplet, (c) edge-wise shared partners}\label{fig_triangle}
\end{figure}

\noindent Empirical studies operationalize triadic closure in different ways.
As mentioned, in undirected binary networks triadic closure can be measured by the percentage of triangles in the network
(see Figure \ref{fig_triangle}a). 
In directed networks, different constellations of incoming and outgoing edges give rise to different interpretations of triadic relationships. 
The most commonly used closed triad in directed networks is the transitive triplet, which indicates that if one node $a$ has two friends $b$ and $c$, then either $b$ ties to $c$ or $c$ ties to $b$ (see Figure \ref{fig_triangle}b).

Inferential network models, such as Exponential Random Graph Models, Latent Space Models or Stochastic Actor-Oriented Models are used to statistically test whether a social network exhibits tendencies for triadic closure. Conventional inferential network models build on binary representations of social networks, where edges between nodes are either present ($=1$) or absent ($=0$). 
In inferential models, triadic closure is often operationalized through the concept of shared partner statistics (see Figure \ref{fig_triangle}c). Shared partner statistics are an edge-based measure and calculate for each edge between nodes $a$ and $b$, how many friends (or more generally \textit{shared partners}) they have in common \cite[see for instance][]{hunter2006inference}. Empirical evidence shows that having mutual friends has a cumulative effect on friendship formation and preservation \cite[][]{martin2006persistence, burt1995kinds} and the statistic therefore captures clustering tendencies in social networks more accurately. 

However, this conventional approach to the operationalization of triadic closure through shared partner statistics poses a problem for multi-edge networks.
There, dyads are equipped with an edge count that reports how many times an interaction between the two nodes has been repeated.
The shared partner statistic therefore has to be adapted to incorporate edge counts (or weights) in order to detect meaningful triadic closure in multi-edge networks. Generally, this raises the question of how the different edge counts can be interpreted. 

Repeated interactions are often considered an indicator of the strength of the relation linking two nodes. 
Reference \citep[][p. 133]{homans1950human} postulates that ``the more frequently persons interact with one another, the stronger their sentiments of friendship for one another are apt to be''. However, the meaning behind repeated interactions are by no means universal and depend on the context of the network as well as the type of interactions used to define edges \cite[][]{marsden1984measuring}.
In friendship networks, repeated interactions in the form of increased communication, interacting between classes or attending social events together outside school potentially indicate strength of friendship interactions. 
However, if the social interaction measures the time and instances of working on schoolwork together it does not automatically translate into friendship relationships; rather the increased interactions could reflect tutelage ties. 
This illustrates that the definition of edges plays an important role in multi-edge networks and in turn, in weighted shared partner statistics.
If repeated interactions are meaningful and reflect edge strength, this strength should be adequately reflected in the measurement of triadic closure. 

\section{Methods}

\subsection{Weighted Shared Partner Statistics for Multi-Edge Networks}

\noindent We propose measuring triadic closure in multi-edge networks using an edge-based approach. For each dyad $(a, b)$ in network $N$, we calculate whether a two-path with dyads $(a, i)$ and $(b, i)$ exist; i.e., we calculate whether nodes $a$ and $b$ have shared partners $i$.
It is important to note that regardless of whether or not nodes $a$ and $b$ have interacted (repeatedly), we calculate whether this dyad $(a, b)$ could potentially close a triad. 
In the ERGM-family, this approach is summarized in so-called change statistics (or change scores) \cite[][]{snijders2006new, hunter2008goodness, krivitsky2011adjusting}. 
The resultant matrix holds the values of the endogenous network statistic  for each dyad and can capture complex patterns between nodes in a network without including the state of the focal dyad $(a, b)$.
This independence of the dyad state $(a, b)$ allows to test whether the values of the pattern correlate with the values of the (repeated) edges in the multi-edge network.

Since the edge counts $v(a,i)$, $v(b,i)$ give the number of edges for each two-path $(a, i)$-$(b, i)$, we incorporate them into the shared partner statistic.
Failing to do so would treat triangles with equal numbers of edge counts and triangles with very different numbers of edge counts the same, which is precisely what should be avoided.
Furthermore, without accounting for edge counts multi-edge networks with high density would result in a change statistic where the very many dyads that show only one edge would be treated as important as dyads with large edge counts. 
By incorporating the edge counts, the variance for the shared partner statistics becomes broader, which allows for more accurate parameter estimates of the effects of triadic closure on the network structure.

\begin{figure*}
    \centering
    \begin{minipage}[b]{0.22\textwidth}
        \centering
        \includegraphics[width=\textwidth]{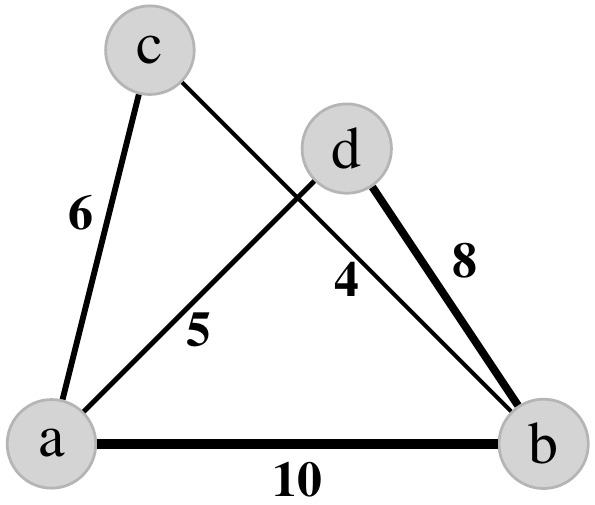} 
	 \caption{Example multi-edge network}\label{fig_examplenw}
    \end{minipage}
    \hfill
    \begin{minipage}[b]{0.22\textwidth}
        \centering
        \captionsetup{type=table} \begin{tabularx}{1\textwidth}{X|XXXX}
		 \toprule
	 		& a & b & c & d \\
		 \midrule
		a	& - & 10 & 6 & 5 \\
		b	& 10 & - & 4 & 8 \\	
		c	& 6 & 4 & - & 0 \\	
		d	& 5 & 8 & 0 & - \\	
		\bottomrule
	 \end{tabularx}
	 \caption{Edge counts $v(i,j)$ of example network}\label{tab_exampleadj}
    \end{minipage}
    \hfill
    \begin{minipage}[b]{0.22\textwidth}
        \centering
        \captionsetup{type=table} 
        \	\centering
	 \begin{tabularx}{1\textwidth}{X|XXXX}
		 \toprule
	 		& a & b & c & d \\
		 \midrule
		a	& - & 2 & 1 & 1 \\
		b	& 2 & - & 1 & 1 \\	
		c	& 1 & 1 & - & 2 \\	
		d	& 1 & 1 & 2 & - \\	
		\bottomrule
	 \end{tabularx}
	 \caption{Shared partner statistic \protect\phantom{weighted}}\label{tab_exampleSP}
    \end{minipage}
    \hfill
    \begin{minipage}[b]{0.22\textwidth}
        \centering
        \captionsetup{type=table} 
        \	\centering
	 \begin{tabularx}{1\textwidth}{X|XXXX}
		 \toprule
	 		& a & b & c & d \\
		 \midrule
 		a	& - & 9 & 4 & 8 \\
 		b	& 9 & - & 6 & 5 \\	
 		c	& 4 & 6 & - & 9 \\	
 		d	& 8 & 5 & 9 & - \\
		\bottomrule
	 \end{tabularx}
	 \caption{Weighted shared partner statistic}\label{tab_exampleWSP}
    \end{minipage}
\end{figure*}

Figure \ref{fig_examplenw} shows a simple multi-edge network with four nodes and Table \ref{tab_exampleadj} reports its corresponding adjacency matrix.
Tables \ref{tab_exampleSP} and \ref{tab_exampleWSP} report binary and weighted shared partner statistics, respectively.
For each dyad in the network, the number of shared partners for said dyad is counted and noted in the unweighted change statistic matrix.
Instead of counting shared partners, the weighted statistic weights each shared partner by the minimum edge count in the two-path $(a,i)$-$(b,i)$. For instance, nodes $a$ and $c$ have one shared partner $b$. The dyads $(a, b)$ and $(c, b)$ have counts 10 and 4 respectively, resulting in a weighted shared partner statistic for dyad ($a, c$) of 4 ($min(v(a,b), v(c,b))$).

Since the example network is nearly complete (only dyad $(c, d)$ is absent), the unweighted shared partner statistic shows little variance compared to the weighted statistic, which accounts for edge intensities. 
When regressing both shared partner statistics against the edge counts of the multi-edge network (i.e., Table \ref{tab_exampleadj}), the unweighted statistic can explain 1\% of the variance, the weighted statistic 4\%\footnote{Estimates obtained as the $R^2$ of the linear regressions network $\sim$ unweighted statistic and network $\sim$ weighted statistic, respectively.}, indicating that more information is stored in the weighted shared partner statistic.

It is important to note that the statistics are dyad-state-independent, i.e., they do not depend on the the edge count of the focal dyad. For instance, nodes $c$ and $d$ do not interact in the example network. However, both nodes have two shared partners in common. The change statistic in its unweighted and weighted form reflects this potential for dyad ($c, d$) to close two triangles. 
Because of this dyad-state-independence, these weighted shared partner statistics can be used as explanatory variables in inferential network models to account for triadic closure. 
We use ERGMs for count data and gHypEG to assess the degree to which social networks are structured by triadic closure. The subsequent paragraphs give a brief introduction to both inferential network models, which is followed by a validation test for our operationalization of triadic closure in multi-edge networks.

\subsection{ERGM} 

\noindent Exponential random graph models (ERGMs) are generative models developed to estimate effects of endogenous and exogenous covariates on network formation \citep[][]{frank1986markov, wasserman1996logit, robins2007introduction, robins2007recent}. The endogenous covariates typically include features of the network, such as degree distributions or triadic closure effects and exogenous covariates are used to estimate homophily effects or effects of nodal attributes on the activity or popularity of nodes \cite[][]{morris2008specification}.

The model can be expressed as the probability of observing the given network $N$ over all possible permutations $\mathcal{N}$ of the network:
\begin{equation}
P(N, \pmb{\theta}) = \frac{\exp\{\pmb{\theta}^T \pmb{h}(N) \}}{\sum_{N^* \in \mathcal{N}} \exp\{\pmb{\theta}^T \pmb{h}(N^*)\}},
\label{ergm}
\end{equation}
where $N$ is the observed network, $\pmb{\theta}$ represent the estimated parameters and $\pmb{h}(N)$ is a vector of statistics containing exogenous and endogenous covariates \citep{cranmer2011inferential}, see also \citep{hunter2006inference}.

Reference \citep[][p. 1105]{krivitsky2012exponential} expanded the ERGM to fit count data by allowing counts to populate dyads (i, j)  in $\mathcal{N}$: 
\begin{equation}
P(N, \pmb{\theta}) = \frac{\pmb{l}(N) \exp\{\pmb{\theta}^T \pmb{h}(N) \}}{\sum_{N^* \in \mathcal{N}}  \pmb{l}(N) \exp\{\pmb{\theta}^T \pmb{h}(N^*)\}},
\label{ergm_count}
\end{equation}
where $\pmb{l}(N)$ sets the shape of the dyad distributions, for instance modeling a poisson distribution. In the binary case of the ERGM, equation \ref{ergm} may reduce to a logistic regression on edge formation \cite[for additional details, see][]{wasserman1996logit, desmarais2012micro}. 
In the count version of the ERGM, equation \ref{ergm_count} may reduce to a Poisson regression model, depending on how $\pmb{l}(N)$ is defined \cite[][]{krivitsky2012exponential}.

Estimation of parameter values in an ERGM is carried out with Markov Chain Monte Carlo Maximum Likelihood Estimation (MCMC MLE) \cite[][]{wasserman1996logit, cranmer2011inferential}. We use the \texttt{ergm.count} package \cite[][]{krivitsky2018ergmcount} in the \texttt{statnet} suit of packages \cite[][]{handcock2003statnet} for the statistical computing environment \texttt{R} to estimate the count ERGMs.

\subsection{GHypEG}

\noindent Generalized hypergeometric ensembles of random graphs (gHypEG) are network models specifically developed to deal with multi-edge networks~(\cite{casiraghi2016generalized, casiraghi2018generalised}).
Similarly to ERGM, they can be used to estimate effects of endogenous and exogenous covariates on network formation \citep{casiraghi2017multiplex}).
They are a generalization of the configuration model (\cite{fosdick2018configuring, molloy1995critical}), where edges are rewired randomly preserving the degree sequences.
Specifically, gHypEG allow to extend the configuration model by specifying the relative propensity of two nodes to be connected that goes beyond what prescribed by their degrees.

The model can be expressed as the probability of observing the given network $N$ by sampling its edges from an urn containing all possible combinations of edges  \citep[][]{casiraghi2018generalised}.
By specifying the number of possible combinations of edges between each pair of nodes $\Xi_{ij} = k^\text{out}_i\cdot k^\text{in}_j$ as a function of their in and out degrees $k^\text{out/in}_i$, the probability for a directed network is given as follows:
\begin{equation}
P(N, \pmb{\theta}) = \prod_{ij} {\Xi_{ij} \choose A_{ij}} \int_0^1\prod_{ij}(1-z^{\frac{\Omega_{ij}}{S_{\pmb{\Omega}}}})^{A_{ij}}dz,
\label{nrm}
\end{equation}
where $S_{\pmb{\Omega}}=\sum_{ij}\Omega_{ij}(\Xi_{ij}-A_{ij})$.

In Eq.\ref{nrm}, $\pmb A$ is the adjacency matrix of the observed network $N$, and $\Omega_{ij} = \exp(\pmb\theta^T\log(\pmb h_{ij}(N)))$ is the relative propensity of two nodes $i,j$ to be connected in terms of the estimated parameters $\pmb{\theta}$ and the vector of statistics $\pmb h_{ij}(N)$ containing exogenous and endogenous dyadic covariates \citep{casiraghi2017multiplex}.
Because of the formulation of gHypEG, the parameters $\pmb{\theta}$ estimate the ``degree-corrected" effect of the covariates as a propensity to connect two nodes beyond what prescribed by degrees.
In the case that such degree-correction should \textit{not} be accounted for, the matrix $\pmb\Xi$ can be constructed according to the average degree $<k>$, as $\Xi_{ij}=<k>^2$ for all $i,j$.

Differently from ERGM, the estimation of parameter values in a gHypEG is carried out with numerical Maximum Likelihood Estimation (MLE) \cite[][]{casiraghi2017multiplex}, thus forgoing computationally intense simulations and allowing the analysis of large networks.
We use the \texttt{ghypernets} package \cite[][]{casiraghi2019ghypernet} to estimate the gHypEGs.

\section{Validation with Synthetic Data}

\noindent We validate our approach to quantifying triadic closure in multi-edge networks by simulating three different multi-edge networks:
The first network is a small network (34 nodes) made up of $1,000$ randomly sampled edges. 
The second network has the same number of nodes but the $1,000$ edges are assigned to a set of randomly selected triangles ($n_{tri} = 26$).
Both networks represent two extremes and do not reflect the structure of real-world social networks.
Rather, they report instances of networks, where triadic closure is the driving factor (example 2) and where triadic closure is present but meaningless (example 1).
The third network represents a combination of the first and second network (34 nodes, 2000 edges). It captures a maximally dense network where half of the edges only form triangles. 
We use this network to test if the weighted change statistic can capture meaningful triadic closure effects in dense multi-edge networks.

\begin{figure}
    \centering
    \subfigure[]{\includegraphics[width=0.33\textwidth]{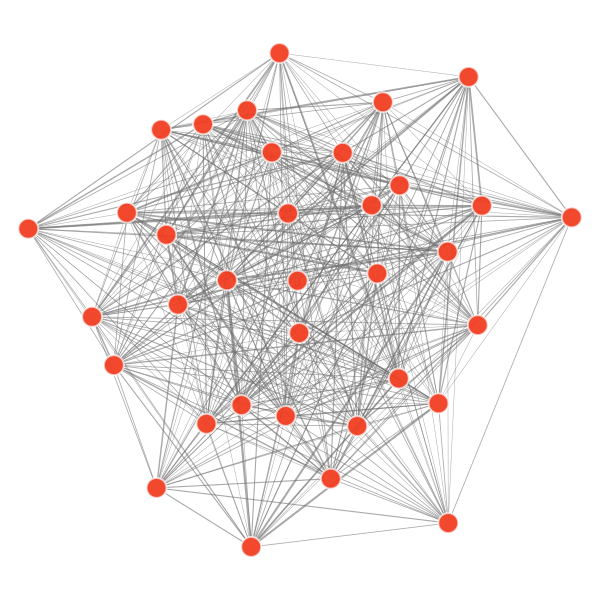}\label{fig_random}}\hfill
    \subfigure[]{\includegraphics[width=0.33\textwidth]{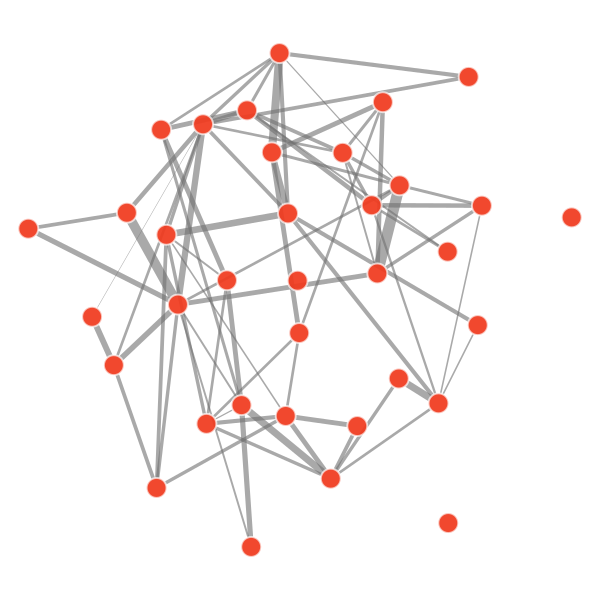}\label{fig_tri}}\hfill
    \subfigure[]{\includegraphics[width=0.33\textwidth]{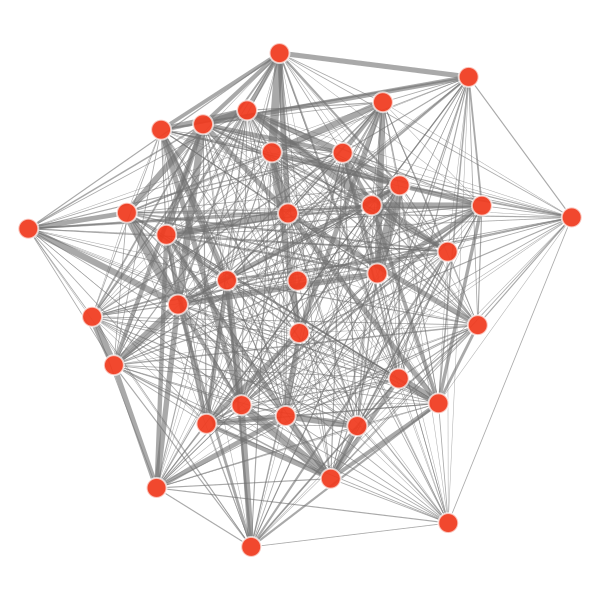}\label{fig_both}} 
    \caption{Three synthetic networks: (a) fully connected with random pairwise edge counts, (b) only repeated triadic interactions, (c) mixture of dyadic and triadic interactions. The underlying topology is fully connected in the (left) and (right).}\label{fig_synthetic}
\end{figure}

\renewcommand{\arraystretch}{1.2}
\begin{table}
\centering
\footnotesize
\begin{tabularx}{\columnwidth}{lXp{.65cm}|Xp{.65cm}|Xp{.65cm}}
\toprule
	 & \multicolumn{2}{c}{(a) random} & \multicolumn{2}{c}{(b) triangles} & \multicolumn{2}{c}{(c) both} \\
	\midrule
	\multicolumn{3}{l}{\textbf{ERGM}}\\
	triadic closure 	& $\phantom{-}0.001$ 	& ($.002$) & $\phantom{-}0.03^{*}$ & ($.002$) & $\phantom{-}0.025^{*}$ & ($.002$)\\
	nonzero			& $-0.07$				& ($0.14$) & $-1.67^{*}$ 		   & ($0.16$) & $-1.67^{*}$			   & ($0.16$)\\
	sum				& $\phantom{-}0.53^{*}$	& ($0.12$) & $\phantom{-}0.13$	   & ($0.09$) & $\phantom{-}0.13$	   & ($0.09$)\\
	\midrule
	\multicolumn{3}{l}{\textbf{gHypEG}}\\
	triadic closure 	& $\phantom{-}0.30$ 	& ($0.19$) & $\phantom{-}1.13^{*}$ & ($0.04$) & $\phantom{-}1.76^{*}$  & ($0.09$)\\
	\bottomrule
	\multicolumn{7}{l}{\scriptsize Coefficients are reported as log-odds; $^*$ indicates p-value $> 0.001$}\\	
\end{tabularx}
\caption{Results of the ERGM and gHypEG on the three synthetic multi-edge networks presented in Figure \ref{fig_synthetic}. Both models report positive and significant results for triadic closure in networks (b) and (c), where edges belonging to 26 random triangles were sampled.}\label{tab_synthetic}
\end{table}

Figure \ref{fig_random} depicts the random network with 34 nodes and 1,000 randomly assigned edges. The multi-edge network has a density of 1, indicating that each node ties at least once with each other node.
This random complete network is thus maximally clustered, i.e., consists of a maximum number of triangles.  
However, these triangles do not explain the structure of the network per se and a inferential network model for multi-edge networks should be able to make a distinction between clustering due to density or clustering due to meaningful triadic closure.
Table \ref{tab_synthetic}, column (a) reports the results of both inferential network models: 
Both models show a near-zero, non-significant effect of triadic closure in the random network.
The results are stable across 1,000 random generations of the synthetic networks, with the gHypEG reporting an average coefficient of -0.09 ($min = -1.55$, $max = 0.73$, $sd = .34$).

Figure \ref{fig_tri} depicts the synthetic network with the triangular structure. We generated a network with 34 nodes and sampled 26 triangles. We then re-sampled 1,000 edges on top of the 26 triangles, thus only re-enforcing the triangular structure in the data.
The results of the count-ERGM and the gHypEG in Table \ref{tab_synthetic}, column (b), confirm that the weighted shared partner statistic significantly correlates with the dyadic counts of the multi-edge network.
The results are stable across 1,000 random generations of the synthetic networks, with the gHypEG reporting an average coefficient of 1.16 ($min = .88$, $max = 1.55$, $sd = .09$).
Since the synthetic network is not complete, the unweighted shared partner statistic (used for binary networks) is able to capture the triadic closure effect as well (gHypEG, $\text{coef} = 1.74$, $SE = 0.05$, $p-value < 0.00$).

Figure \ref{fig_both} depicts the synthetic network made up of random edges (Network (a)) and randomly sampled triangles (Network (b)).
Table \ref{tab_synthetic}, column (c), reports the results of both inferential network models. 
Both models report a positive and highly significant effect of triadic closure for Network (c), as expected.
The results are stable across 1,000 random generations of the synthetic networks, with the gHypEG reporting an average coefficient of 1.81 ($min = 0.94$, $max = 2.44$, $sd = 0.20$).

However, when calculating the unweighted shared partner statistic, the triadic closure effect cannot be captured by the gHypEG (nor the ERGM). 
The gHypEG reports a near-zero and non-significant coefficient $\beta = 0.08$ ($SE = 0.20$, $p-value = 0.694$).
Conventional approaches to measuring triadic closure (as used for binary networks) fail here, because the small network of 34 nodes with 2,000 randomly drawn edges is too dense to distinguish meaningful triadic closure.
By neglecting edge counts, triadic closure in complete (or near complete) networks cannot be adequately estimated or controlled for. 
The weighted shared partner statistic is necessary to estimate whether triadic closure is a meaningful driving force of the network structure.

\section{Case Studies}

\begin{figure*}
    \centering
    \begin{minipage}[b]{0.33\textwidth}
        \centering
        \includegraphics[angle=90,width=.6\textwidth]{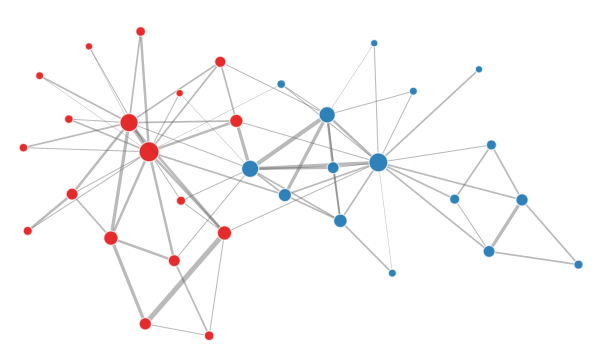}
        \caption{Zachary’s Karate Club network. Nodes are colored by factions.}\label{fig_karate}
	\end{minipage}
	\hfill	
    \begin{minipage}[b]{0.65\textwidth}
		\footnotesize
		\centering
		\captionsetup{type=table}
		\begin{tabular}{@{}lrlc|rlc@{}}
			\toprule
				& \multicolumn{3}{c}{\textbf{ERGM}} & \multicolumn{3}{c}{\textbf{gHypEG}} \\
				\midrule
				nonzero 	& 	$-$3.281	& $^{***}$ 		& (0.267) 	& \qquad--&	& \\ 
				sum 		& 	$-$1.166	& $^{***}$ 		& (0.297) 	& \qquad--&	& \\ 
				degree dist. 	& 	0.028	& $^{***}$ 	& (0.004) 	& \qquad--&	& \\ 
				triadic closure & $-$0.016	&	 		& (0.012)	& $-$0.160 & $.$ 	& (0.086) \\ 
				faction 		& 	1.123	& $^{***}$ 		& (0.178)	& ~~1.090 & $^{***}$ 	& (0.104) \\ 
					\midrule
				AIC 		& 	\phantom{47,}$-$869.4& 	&	 &	\phantom{1,34,}674.7 & 	&	\\
				Null AIC 	& 	0		& 	&	 &	869.1 & 	&	\\ 
				\bottomrule
		\end{tabular}
		\vspace{0.5em}
\caption{ERGM and gHypEG results on the Zachary's Karate Club network.}\label{tab_karate} 
    \end{minipage}
	\vspace{1em}

    \begin{minipage}[b]{0.33\textwidth}
        \centering
        \includegraphics[width=1\textwidth]{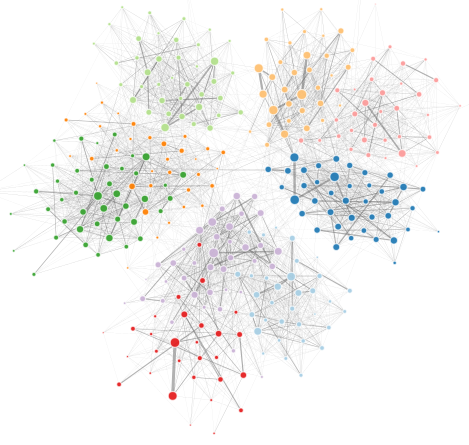}
        \caption{Friendship network of high school students colored by classes.}\label{fig_friends}
	\end{minipage}
	\hfill
    \begin{minipage}[b]{0.65\textwidth}
		\footnotesize
		\centering
		\captionsetup{type=table}
		\begin{tabular}{@{}lrlc|rlc@{}}
			\toprule
				& \multicolumn{3}{c}{\textbf{ERGM}} & \multicolumn{3}{c}{\textbf{gHypEG}} \\
				\midrule
				nonzero 		& $-$4.033 & $^{***}$  		& (0.006) & \qquad--&	& \\ 
				sum 			& 0.239 & $^{***}$ 			& (0.004) & \qquad--&	& \\ 
				degree dist. 	& 0.00002 & $^{***}$  	& (0.000) & \qquad--&	& \\ 
				triadic closure & 0.008 & $^{***}$ 		& (0.000) & 0.819 & $^{***}$ & (0.003) \\ 
				class 			& 0.162 & $^{***}$ 		& (0.004) & 0.879 & $^{***}$ & (0.005) \\ 
					  & &\\ 
			   \midrule
			   AIC 	& $-$47,511.5	& & & 593,853.5 \\
			   Null AIC 	& 0 	& & & 1,346,750\\ 
			   \bottomrule
		\end{tabular}
		\vspace{3em}

\caption{ERGM and gHypEG results on a high school friendship network.} 
  \label{tab_friend} 

    \end{minipage}
\end{figure*}

\subsection{Karate Club Network}

\noindent In our first case study we examine whether the interaction network of 34 karate club members at a U.S. University reported by \citep[][]{zachary1977information} show signs of triadic closure.\footnote{The dataset is widely known under the name \textit{Zachary’s Karate Club dataset}.} 
Reference \citep[][]{zachary1977information} recorded different social contexts in which 34 members interacted, for instance attending a university bar, participating in the same tournament or attending a weekend karate course.
Out of these different interactions, a network with 34 nodes was constructed, counting for each dyad $(i, j)$ how often two nodes $i$ and $j$ were part of the same social context.

The data set has a known block structure. 
During the course of the study the members split into two factions, lead by two central nodes \cite[][]{zachary1977information}.
Figure \ref{fig_karate} depicts the network with the 34 nodes colored by their faction.
In addition to triadic closure, we therefore control for this known block structure in the network.
Furthermore, we control for degree distributions to examine whether triadic closure is a simple artifact of the degree distribution. 
In the karate club data, (repeated) social interactions have a very specific meaning.
Instead of accounting for direct interactions among members, the co-occurrence by two members in the same social contexts are recorded.
We therefore assume that triadic closure does \textit{not} play a strong role in this network.
It is fairly unlikely, that if member $a$ attends an academic class with $b$, and $b$ visits a weekend karate lesson with $c$, $a$ and $c$ interact in a third context (e.g., at a bar). 
Such a closing triad is only plausible, if co-occurrence in the same social contexts also means direct social interactions among all members present.

The results of the two inferential network models are reported in Table \ref{tab_karate}.
Both models report a negative, non-significant coefficient of triadic closure. 
Holding node degrees and faction membership constant, triadic closure does not add to the explanation of the network structure.
As stated above, the nature of the edge counts in the Karate Club network is based on co-occurrence in the same social contexts. As they do not measure direct social interactions it is highly plausible that triadic closure does not explain co-occurrence in social contexts.
Furthermore, the network is divided into two strong factions with minimal interactions between the two factions. 
This is another indicator of weak closure in the network. 
There may be closure within the factions, but the strong division does not give evidence of closure over the full network.
Both models report a positive and significant effect of faction homophily. Nodes from one faction are more likely to share the same social contexts with nodes from the same faction.
For the count ERGM, three additional model terms are reported. 
The \texttt{sum}-term represents an intercept term for edge counts. 
It controls for the expected number of interactions (i.e., edge counts) in the multi-edge network.
The \texttt{nonzero}-term reports a negative and significant coefficient, indicating that there is some zero-inflation in the data. 
This is due to the two faction leaders acting like stars in the network and helping to split the network into two factions.
We also control for degree distributions in the ERGM count network to ensure that the triadic closure effect is not masked by degree-based effects. 
The gHypEG already reports degree-corrected parameters.

\subsection{Friendship Network in a French High School}

\noindent In our second case study we examine a multi-edge friendship network, reported by \citep[][]{mastrandrea2015contact}. 
The network consists of 327 nodes, each indicating a student attending a French high school, and 188,508 undirected ties, measured using contact sensors.
For each dyad, we know whether or not they attended classes together.
Classroom homophily is an important factor in friendship networks as they provide increased opportunity to meet. 
Figure \ref{fig_friends} depicts the network. 
Nodes are colored by school classes and show increased friendship interaction within classes than between.
We therefore include the information of school classes in the inference models for a more severe test of triadic closure in this network.
Triadic closure has been theorized to be the product of increased opportunity to interact \cite[see e.g.,][]{simmel1950sociology}.
Attending classes together increases the opportunity for students to interact and become friends, by controlling for this potential exogenous influencing factor, we can test whether triadic closure drives friendship formation patterns beyond class-based interactions.

Results of the two inferential network models are reported in Table \ref{tab_friend}.
Both models report a positive and significant effect of triadic closure in these multi-edge friendship networks. 
Triadic closure has been widely studied in friendship networks \cite[see e.g.,][]{rambaran2015development, mercken2010dynamics} and is commonly known to be an important driving factor of friendship formation.
Both models confirm these previous studies.
It is important to note that both models correct for degree distributions and that triadic closure in this friendship network goes beyond degree-based clustering.

\section{Discussion}

\noindent Conventional approaches to the operationalization of endogenous network statistics fail if used on non-binary networks. 
These statistics were developed for binary networks and do not easily translate into multi-edge, weighted or valued networks.
Built on theoretical ideas of closure in social networks and empirical evidence thereof, we expand a measure of triadic closure in social networks to multi-edge networks of social interactions.

Social networks and the relations they capture between their entities are rarely binary in nature.
The relations and interactions that make up the edges in social networks are part of a complex, temporal process.
Friendship relations are formed through repeated interactions of supportive
nature. 
Once formed, these friendship relations are reinforced and maintained through repeated interactions and support.
However, social network analysis has in the past abstracted these social dynamics into static snapshots with edges being either present or absent.
Binary networks represent a simplification that is often made to accommodate standard tools of social network analysis. 
This simplification introduces strong biases on the results of the analysis due to considerable loss of information on the relations \cite[see][]{thomas2018valued}.

However, with increased information on social interactions and better techniques of measuring and storing said information, social network analysis can move beyond the analysis of static, binary networks.
This entails updating standard network analysis concepts to incorporate additional information, either on the temporal process of edge formation or the strength of edges. 
In this paper, we present a new operationalization of triadic closure for social networks that contains information on repeated interactions among nodes.
Our approach is based on theoretic ideas of closure and empirical evidence.
We operationalize triadic closure using an edge-based approach in the form of a shared partner statistic.
We incorporate repeated interactions into our measure by calculating for each dyad, the minimum number of interactions two nodes $a$ and $b$ should have in order to close a triad with a third node $c$.
Furthermore and based on empirical evidence of the cumulative effect of shared partners on edge formation \cite[e.g.,][]{martin2006persistence}, our measure sums minimal edge counts over all shared partners of a particular dyad.

Our operationalization of triadic closure is able to detect meaningful closure in multi-edge networks where standard operationalizations of triadic closure fail.
Using synthetic data, we show that even in dense multi-edge networks, meaningful triadic closure can be detected and controlled for.
Our operationalization allows for inferential network models of multi-edge networks, such as the Exponential Random Graph Model (ERGM) for valued networks or the generalized hypergeometric ensembles of random graphs (gHypEG), to test or control for one of the most important relational mechanisms in social networks.

We validated our operationalization by generating three synthetic multi-edge networks with 34 nodes. 
The first network is made up of 1,000 random edges and both the ERGM and the gHypEG correctly report a non-significant near-zero effect of triadic closure.
The second network is made up of 1,000 edges belonging to 26 triangles and both the ERGM and the gHypEG correctly report positive effects of weighted triadic closure. 
For the third network, we superimpose the first network onto the second to test if triadic closure can be detected even if the network is dense. 
Both models report significant and positive effects of triadic closure, validating our approach.
We further tested our operationalization of triadic closure for multi-edge networks on two real-world data examples.
Our first case study are social interactions of members of a karate club \cite[][]{zachary1977information}. 
Here, both models correctly identify an absence of triadic closure.
Our second case study are proximity contacts between students in a high school \cite[][]{mastrandrea2015contact}.
Here, triadic closure plays an important role in the network and is correctly identified in both inferential models.

Our operationalization of one of the most important relational mechanisms in social networks allows inferential network analyses to move away from the analysis of binary networks towards the analyses of multi-edge and weighted networks, which offer a more realistic representation of social interactions and relations.

\end{document}